\documentclass[twocolumn,trackchanges]{aastex631}

\usepackage{amsmath}
\usepackage{graphicx}
\usepackage{subfigure}

\begin{document}

\title{Morphology of molecular clouds at kiloparsec scale in the Milky Way: Shear-induced alignment and vertical confinement}

\author[0009-0004-5094-3522]{Yi-Heng Xie}
\affiliation{South-Western Institute for Astronomy Research, Yunnan University, Chenggong District, Kunming 650500, P. R. China}

\author[0000-0003-3144-1952]{Guang-Xing Li}
\altaffiliation{E-mail: gxli@ynu.edu.cn (G-XL)}
\affiliation{South-Western Institute for Astronomy Research, Yunnan University, Chenggong District, Kunming 650500, P. R. China}

\author[0000-0003-2472-4903]{Bing-Qiu Chen}
\altaffiliation{E-mail: bchen@ynu.edu.cn (B-QC)}
\affiliation{South-Western Institute for Astronomy Research, Yunnan University, Chenggong District, Kunming 650500, P. R. China}



\begin{abstract}
The shape of the cold interstellar molecular gas is determined by several processes, including self-gravity, tidal force, turbulence, magnetic field, and galactic shear. Based on the 3D dust extinction map derived by Vergely et al., we identify a sample of 550 molecular clouds (MCs) within $3\hspace{0.2em}\rm kpc$ of the solar vicinity in the Galactic disk. Our sample contains clouds whose size ranges from pc to kiloparsec, which enables us to study the effect of Galactic-scale processes, such as shear, on cloud evolution. We find that our sample clouds follow a power-law mass-size relation of $M\propto 32.00\hspace{0.2em}{R_{\rm{max}}}^{1.77}$, $M\propto 20.59\hspace{0.2em}{R_S}^{2.04}$ and $M\propto 14.41\hspace{0.2em}{R_V}^{2.29}$, where $R_{\rm{max}}$ is the major axis-based cloud radius, $R_S$ is the area-based radius, and $R_V$ is the volume-based radius, respectively. These clouds have a mean constant surface density of $\sim 7 \hspace{0.2em} \rm {M_{\odot}pc^{-2}}$, and follow a volume density-size relation of $\rho \propto 2.60\hspace{0.2em}{R_{\rm{max}}}^{-0.55}$. As cloud size increases, their shapes gradually transition from ellipsoidal to disk-like to bar-like structures. Large clouds tend to have a pitch angle of $28^{\circ} - 45^{\circ}$, where the angle is measured concerning the Galactic tangential direction. These giant clouds also tend to stay parallel to the Galactic disk plane and are confined within the Galactic molecular gas disk. Our results show that large molecular clouds in the Milky Way can be shaped by Galactic shear and confined in the vertical direction by gravity. 

\end{abstract}



\section{Introduction} 
\label{sec:intro}

The evolution of molecular clouds is a complex process dominated by multiple interactions such as self-gravity (e.g. \citealt{Andr2014, Li2017}), tidal force (e.g. \citealt{Lee2018, Liu2021, Li2024}), turbulence (e.g. \citealt{Larson1981, Elmegreen2004, Scalo2004, Ballesteros2007}), magnetic field (e.g. \citealt{Richard2012, LiHB2014}), and Galactic shear (e.g. \citealt{Li2013, Li2016, Jeffreson2018, Liu2021, Li2022}). In 1980s \citet{Larson1981} found three fundamental relations in molecular clouds which are known as Larson's laws: (1) The velocity dispersion and size of molecular clouds follow a power-law relation of $\sigma_v\sim R^\beta$, where $\beta=0.38$ (recent studies have derived a slightly larger exponent of $\beta \sim 0.5-0.6$, e.g. \citealt{Solomon1987, Bolatto2008, Miville2017}). (2) The molecular clouds are roughly in virial equilibrium (their virial parameters $\alpha_{\rm vir}\sim 1$). (3) The mass surface densities of clouds are approximately constant. This third law can also be explained as the clouds follow a mass-size relation of $M\sim R^2$, and recent observational results also derived similar conclusions (e.g. \citealt{Roman2010, Miville2017, Chen2020, Lada2020}). 

The morphology is a key to understanding the evolutionary process of molecular clouds. Previous studies showed that filamentary molecular clouds are ubiquitous and varied in the Milky Way, and their spatial structures are irregular and complex with a wide range of size from a few tenths to hundreds pc (e.g. \citealt{Bally1987, Goldsmith2008, Li2013, Wang2015, Zucker2015, Li2016, Hacar2023}). Recent studies have made several catalogs of the morphology of large-sample molecular clouds based on sky surveys (e.g. \citealt{Yuan2021, Neralwar2022a, Neralwar2022b}). Galactic shear is an effect in which the gas at different locations can have different velocities of Keplerian motion. Thus, large coherent structures can be stretched by this differential motion. The preferential direction where clouds are aligned is perpendicular to the angular momentum vector of the gas. Shear-induced alignment has been reported in cloud-sized filamentary structures in the Milky Way \citep{Li2016}, and kpc-sized structures \citep{Li2022}. Simulations have further shown that amounts of large-size filamentary molecular clouds, which are elongated by Galactic rotation, can exist in shear-dominated inter-arm regions (e.g. \citealt{Smith2014, Duarte2017}). 

However, due to the limitations of observational techniques, we still lack a comprehensive picture of the three-dimensional spatial structure of molecular clouds in the Milky Way. Another problem is that most previous studies focused on the clouds from a few tenths pc to hundreds pc, and similar investigations at the kiloparsec scale are still incomplete. The question of how the morphological properties of clouds are affected by the Milky Way itself has not been fully unresolved.  

In this paper, we present an observational study within a $6\times6\times0.8\hspace{0.2em}\rm kpc$ region around the Sun in the Galactic plane based on the three-dimensional dust extinction map derived by \citet{Vergely2022}. We identify 550 molecular clouds via the Python program \textsc{Dendrogram} \citep{Rosolowsky2008} and derive their mass-size relation at the kiloparsec scale. We investigate their spatial structures and distinguish them into several types of shapes. We also calculate their orientations and extension along different directions in the Milky Way.



\begin{splitdeluxetable*}{ccccccccccccBccccB}
\tabletypesize{\scriptsize}
\tablewidth{0pt} 
\tablecaption{Catalogue of molecular clouds \label{tab:cloud}}
\tablehead{
\colhead{ID}  &  \colhead{GAL X}  & \colhead{GAL Y}  & \colhead{GAL Z}  & \colhead{l}  & \colhead{b}  & \colhead{d}  & \colhead{Mass}  & \colhead{$R_{\rm max}$}  & \colhead{$R_{\rm mid}$}  & \colhead{$R_{\rm min}$}  & \colhead{Shape} & \colhead{Pitch angle} & \colhead{Angle with Galactic disk}  & \colhead{Extension along $\vec R_{\rm gal}$}  & \colhead{Extension along $\vec z_{\rm gal}$}\\ 
\colhead{} & \colhead{(pc)} & \colhead{(pc)}& \colhead{(pc)} & \colhead{($^{\circ}$)} & \colhead{($^{\circ}$)} & \colhead{(pc)} & \colhead{($M_{\odot}$)} & \colhead{(pc)} & \colhead{(pc)} & \colhead{(pc)} & \colhead{} & \colhead{($^{\circ}$)} & \colhead{($^{\circ}$)} & \colhead{(pc)} & \colhead{(pc)}
} 
\colnumbers
\startdata
1 & -648.55  & -485.04  & -175.74  & 216.79  & -12.24  & 828.71  & 295107.08  & 152.96  & 55.84  & 42.99  & bar & 41.54  & 11.35  & 108.18  & 60.44  \\
2 & -1928.32  & -1400.50  & -106.01  & 215.99  & -2.55  & 2385.59  & 282924.14  & 180.80  & 108.83  & 40.25  & disk & 49.65  & 21.14  & 152.50  & 62.55  \\
3 & -570.03  & -408.06  & -225.85  & 215.60  & -17.86  & 736.52  & 35554.30  & 50.73  & 34.56  & 14.26  & disk & 79.73  & 3.39  & 49.86  & 19.34  \\
4 & -564.86  & -389.93  & -230.56  & 214.62  & -18.57  & 724.06  & 7595.85  & 28.52  & 13.40  & 9.31  & bar & 27.07  & 35.23  & 16.64  & 15.49  \\
5 & -672.57  & -500.74  & -167.35  & 216.67  & -11.29  & 855.04  & 112404.79  & 111.28  & 36.90  & 30.52  & bar & 25.04  & 32.21  & 58.06  & 38.91  \\
6 & -659.40  & -474.08  & -174.40  & 215.71  & -12.12  & 830.65  & 80521.50  & 59.50  & 33.96  & 30.24  & ellipsoid & 8.97  & 25.05  & 34.30  & 30.96  \\
7 & -232.54  & 57.41  & -215.11  & 166.13  & -41.93  & 321.93  & 13157.42  & 59.26  & 21.07  & 16.43  & bar & 53.39  & 21.59  & 49.50  & 24.46  \\
8 & -597.76  & -416.91  & -228.96  & 214.89  & -17.44  & 763.90  & 6343.72  & 17.83  & 13.28  & 9.75  & ellipsoid & 54.81  & 0.52  & 16.53  & 10.98  \\
9 & -657.35  & -454.09  & -176.44  & 214.64  & -12.45  & 818.19  & 21764.62  & 36.11  & 28.51  & 16.09  & disk & 40.23  & 4.92  & 31.81  & 22.25  \\
10 & -269.82  & -1643.91  & -42.69  & 260.68  & -1.47  & 1666.45  & 402603.82  & 200.92  & 78.21  & 59.70  & bar & 19.99  & 36.51  & 98.02  & 77.19  \\
... & ... & ... & ... & ... & ... & ... & ... & ... & ... & ... & ... & ... & ... & ... & ... \\
548 & -268.72  & 920.84  & 232.12  & 106.27  & 13.60  & 986.93  & 4393.70  & 16.87  & 9.90  & 8.97  & ellipsoid & 13.62  & 29.22  & 10.16  & 15.27  \\
549 & -318.86  & 908.23  & 234.85  & 109.35  & 13.71  & 990.81  & 9423.68  & 19.05  & 13.68  & 9.72  & ellipsoid & 37.07  & 23.58  & 15.19  & 11.09  \\
550 & -301.20  & 966.59  & 239.18  & 107.31  & 13.29  & 1040.30  & 2927.56  & 25.54  & 11.34  & 7.04  & bar & 66.63  & 12.74  & 23.49  & 12.55  \\
\enddata
\end{splitdeluxetable*}
  

\section{Data \& Method}

\subsection{Data}

Our molecular cloud structures are identified from the 3D dust extinction map derived by \citet{Vergely2022}. They provided the inter-calibration method to calibrate two different types of extinction data: one is based on both spectroscopy and photometry measurement, and another is based on pure photometry from Gaia Early Data Release 3 (EDR3, \citealt{Gaia2021}) photometric data with Two Micron All-sky Survey measurements (2MASS, \citealt{2MASS2006}). They further applied the hierarchical inversion technique described in \citet{Vergely2010} and \citet{Lallement2019} to the calibrated data. They provided several maps with different spatial resolutions and extents (wide extent corresponds to lower resolution) in the solar vicinity. Based on the balance between resolution and extent, we adopt the one with a volume of $6\times6\times0.8\hspace{0.2em}\rm {kpc^3}$ and a spatial resolution of $5\times5\times5\hspace{0.2em}\rm {pc^3}$ (the magnitude of the map is defined at a wavelength of $\rm {5500\hspace{0.2em}\mathring{A}}$).      

In this map, they used the inversion method to estimate the local differential extinction, which imposes spatial correlations between the volume densities of ISM \citep{Lallement2019}. To smooth the volume opacity data they imposed a minimum spatial size and maximum spatial gradients for the distribution of absorbing dust. The correlation length they used is $25\hspace{0.2em}\rm {pc}$, thus the structures with smaller scales within the map are isotropic by construction \citep{Vergely2022}. However, this map can effectively represent the actual morphology of cloud structures at a larger scale.


\subsection{Structure Identification}

We use the Python program \textsc{Dendrogram} \citep{Rosolowsky2008} to identify cloud structures. \textsc{Dendrogram} is designed to extract the tree structure within the dataset using the topological relation between isosurfaces. As the extinction map contains the mass density information of the interstellar medium automatically (assuming a constant dust-to-gas ratio), we can directly apply the algorithm to the extinction map. \textsc{Dendrogram} has three free parameters: $min\_value$, $min\_delta$ and $min\_npix$: (1) $min\_value$ is the minimum value that is used to neglect any structure below it; (2)$min\_delta$ is used to identify the substructures within the superstructures; (3)$min\_npix$ is the minimum pixel number an individual structure needs. 


\subsection{Structure Parameters}

\subsubsection{Mass}

We use the formula below to calculate the mass of each  cloud structure:
\begin{equation}
{M}=\mu m_{\rm {H}}V(\frac{N_{\rm {H}}}{A_{\rm {V}}})\sum_{i}^{n}A_{\rm {V,i}}
\label{eqn:mass}
\end{equation}
where $\mu$ is the mean molecular weight, $m_H$ is the mass of hydrogen atom, $V$ is the physical volume of one pixel in the 3D extinction map (in our case, $5^3\hspace{0.2em}\rm {pc^3}$), $A_{\rm {V}}/N_{\rm {H}}$ is the ratio of visual extinction value to total hydrogen column density, $A_{\rm {{V,i}}}$ is the optical extinction value for each pixel within the cloud structure in unit of $\rm {mag\hspace{0.2em}pc^{-1}}$, $n$ is the number of pixels each structure contains. In our work, we adopt $\mu=1.37$ \citep{Lombardi2011} and $A_{\rm {V}}/N_{\rm {H}}=4.15\times10^{-22}\hspace{0.2em}\rm {mag\hspace{0.2em}cm^{2}}$ \citep{Chen2015}.

\subsubsection{Morphological Parameters}
\label{Morphological parameter}

We use a moment-of-inertia-based approach to derive the morphological parameters of cloud structures. The sphere center of each structure is determined by its mean position in the 3D map along three coordinate axes ($X_{\rm {mean}}$, $Y_{\rm {mean}}$ and $Z_{\rm {mean}}$). The inertia matrix of each structure is given by:

\begin{equation}
I = \begin{bmatrix}
 \sigma_{xx}^{2} & \sigma_{xy}^{2} & \sigma_{xz}^{2}\\
 \sigma_{xy}^{2} & \sigma_{yy}^{2} & \sigma_{yz}^{2}\\
 \sigma_{xz}^{2} & \sigma_{yz}^{2} & \sigma_{zz}^{2}
\end{bmatrix}
\label{eqn:matrix}
\end{equation}
where:

\begin{equation}
\sigma_{\mu\nu}=\sqrt{\frac{\sum_{i}A_i(\mu_i-\mu_{\rm {mean}})(\nu_i-\nu_{\rm {mean}})}{\sum_{i}A_i}}
\label{eqn:sigma}
\end{equation}
where $\mu, \nu \in (x,y,z)$, and $A_i$ is the extinction value of each pixel within the structure. We use the second-order moment (i.e. three eigenvectors of $I$) as cloud size estimation, and no underlying profiles are assumed. To derive an appropriate ratio between the second-order moment with cloud structure radius, we assume a three-axis ellipsoid cloud with Gaussian density profile as follows:
\begin{equation}
\rho(r_i) = \rho_0\frac{1}{\sqrt{2\pi}\sigma_i}{\rm exp}(-\frac{r_i^2}{2\sigma_i^2})
\label{eqn:Gaussian}
\end{equation}
and its standard deviations $(\sigma_i)$ along three axes are derived from three eigenvalues $(\lambda_{\rm max}>\lambda_{\rm mid}>\lambda_{\rm min})$ of $I$. Then the density-weighted radii of clouds along three axes $(R_{\rm {max}}>R_{\rm {mid}}>R_{\rm {min}})$ are defined as:

\begin{equation}
R_i = \frac{\int_{0}^{\infty}r\cdot4\pi r^2\rho(r_i,\sigma_i)dr}{\int_{0}^{\infty}4\pi r^2\rho(r_i,\sigma_i)dr} \approx 1.6\,\lambda_i
\label{eqn:radius}
\end{equation}

The area of clouds is defined as the area of the ellipse with two semiaxes $R_{\rm {max}}$ and $R_{\rm {mid}}$, and the volume of clouds is defined as the volume of the ellipsoid with three semiaxes $R_{\rm {max}}$, $R_{\rm {mid}}$ and $R_{\rm {min}}$ as follows:

\begin{equation}
S=\pi R_{\rm {max}}R_{\rm {mid}}
\label{eqn:area}
\end{equation}

\begin{equation}
V=\frac{4}{3}\pi R_{\rm {max}}R_{\rm {mid}}R_{\rm {min}}
\label{eqn:volume}
\end{equation}

Based on the definitions above, we can define the mass surface density $(\Sigma)$ and the mass volume density $(\rho)$ of clouds as follows:

\begin{equation}
\Sigma = \frac{M}{S}
\label{eqn: surface density}
\end{equation}

\begin{equation}
\rho = \frac{M}{V}
\label{eqn: volume density}
\end{equation}


\begin{figure*}
\includegraphics[width=6.6in]{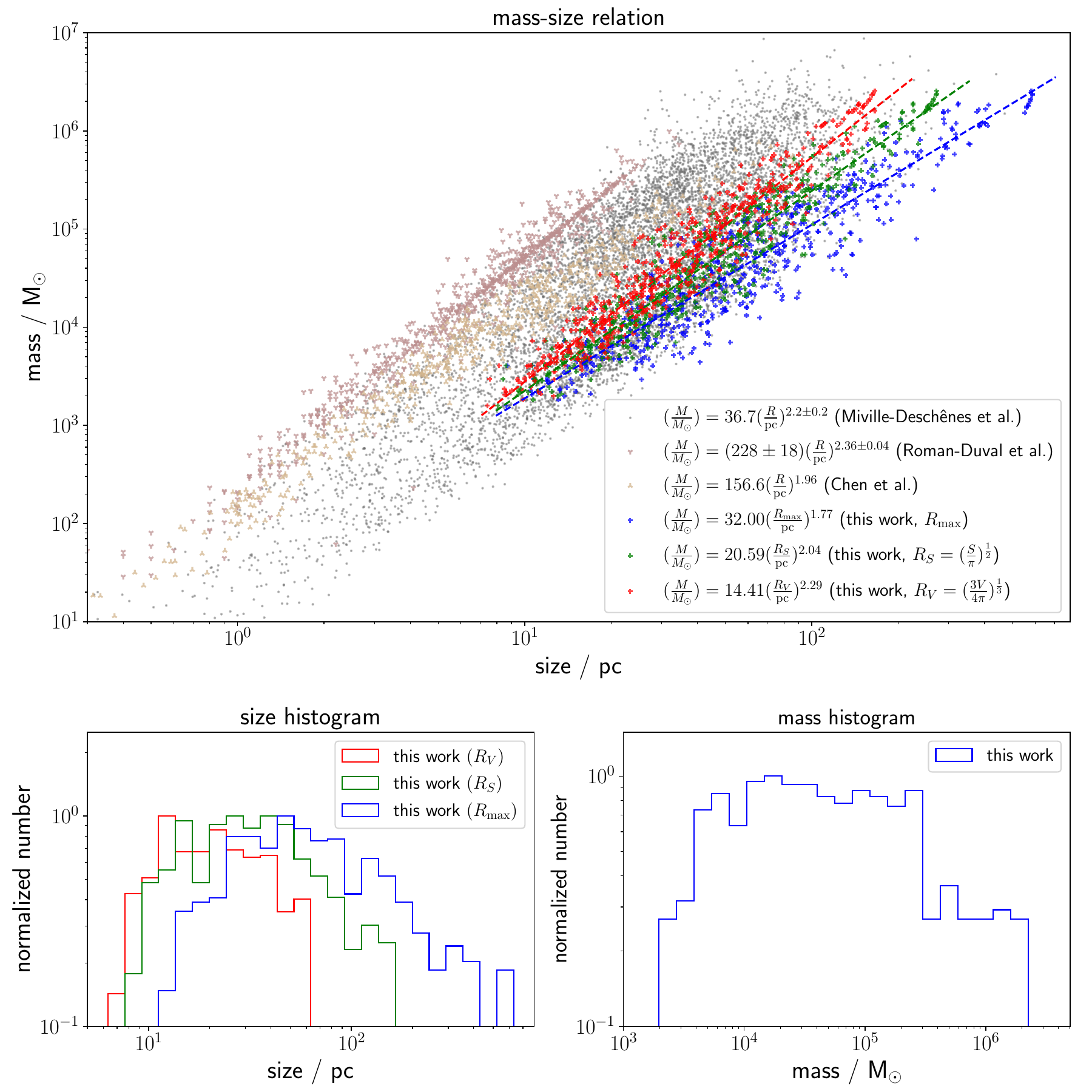}
\centering
\caption{\textbf{The mass-size relations and histograms of clouds.} (1) Upper panel: mass-size relations from our work in three different cloud size definitions: semi-major axis-based definition (blue scatters), area-based definition (green scatters) and volume-based definition (red scatters), and three previous works (\citealt{Miville2017}, gray scatters; \citealt{Roman2010}, rosy-brown scatters; \citealt{Chen2020}, tan scatters). The fitting of our results are $M\propto 32.00\hspace{0.2em}{R_{\rm{max}}}^{1.77}$ (blue dashed line), $M\propto 20.59\hspace{0.2em}{R_S}^{2.04}$ (green dashed line), and $M\propto 14.41\hspace{0.2em}{R_V}^{2.29}$ (red dashed line) respectively. (2)Lower panel: histograms of three different definition-based cloud sizes (left panel) and cloud masses (right panel) from our work.}
\label{fig:m-s}
\end{figure*}

\begin{figure*}
\centering
\includegraphics[width=\textwidth]{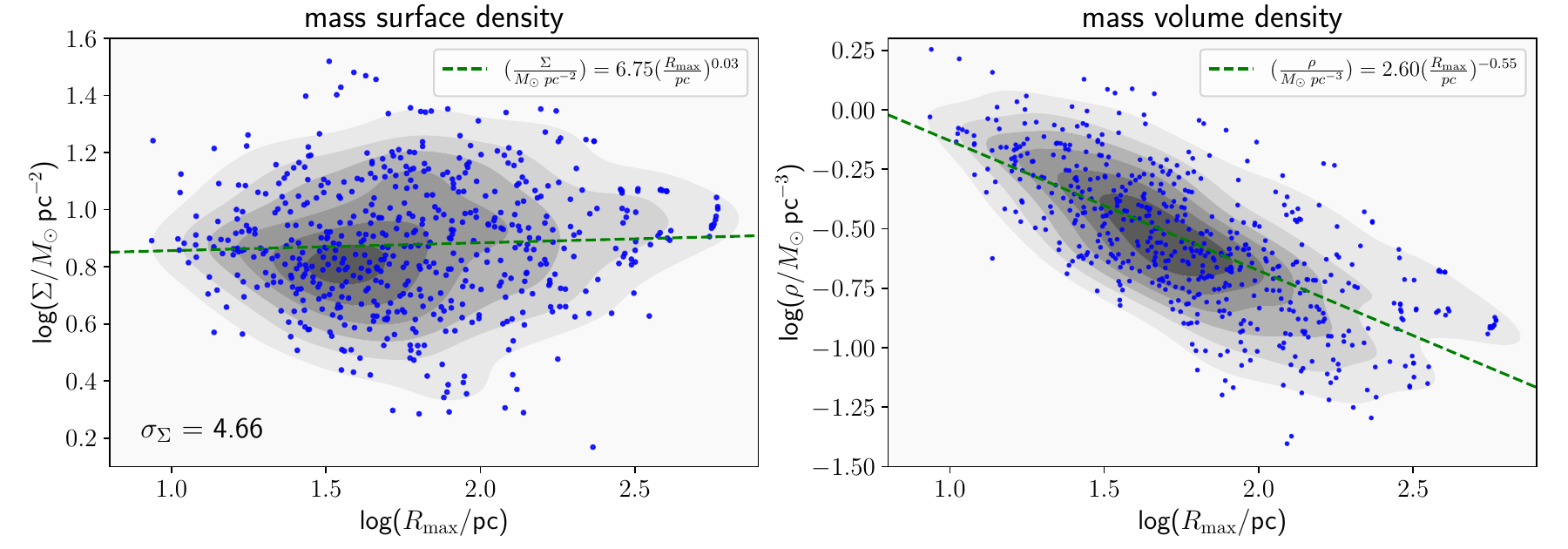}
\caption{\textbf{Mass surface density and mass volume density of clouds.} (1) Left panel: Surface density $(\Sigma)$ of sample clouds versus cloud size. The fitting result is $\Sigma \sim 6.75{R_{\rm{max}}}^{0.03}$ (green dashed line). The standard deviation of the cloud surface density is $\sigma_{\Sigma} = 4.66$. (2) Volume density $(\rho)$ of sample clouds versus cloud size. The fitting result is $\rho \sim 2.60{R_{\rm{max}}}^{-0.55}$ (green dashed line). The background gray colormap shows the result of kernel density estimation.}
\label{fig:density}
\end{figure*}

\section{Result \& Discussion}

To derive a suitable size of molecular cloud structures using \textsc{Dendrogram}, we set $min\_value = 0.002\hspace{0.2em}\rm{mag\hspace{0.2em}pc^{-1}}$, $min\_delta = 0.003\hspace{0.2em}\rm{mag\hspace{0.2em}pc^{-1}}$, and $min\_npix = 20$ through experiment. This configuration corresponds to the threshold of volume density about $5.29\times 10^{-2}\hspace{0.2em}\rm {M_{\odot}pc^{-3}}$ (or $n_{\rm {H}}\approx2.14\hspace{0.2em}\rm {cm^{-3}}$). Based on the identified catalog, we exclude some structures that contain several independent substructures or are too close to the boundary of the extinction map. We finally obtained 550 cloud structures within the 3D dust map. Their physical and morphological properties are partly listed in Table.\ref{tab:cloud} and discussed below.

\subsection{Mass-Size Relation}
\label{m-s}

Our sample molecular clouds exhibit a mass range spanning over three orders of magnitude ($\sim10^3-3\times10^{6}$ $\rm {M_{\odot}}$), and a size range covering about two orders of magnitude ($\sim10^1-10^3 \rm {pc}$). These clouds have a line mass ($M/2R_{\rm{max}}$) range of $\sim 50-3000\hspace{0.2em}\rm {M_{\odot}pc^{-1}}$. The largest structure in our sample has an approximate semi-major axis of $600\hspace{0.2em}\rm {pc}$ and contains more than $2.5\times10^6$ solar mass. Using this \textsc{Dendrogram} approach, the clouds are defined as spatially coherent entities, where no dynamical stability is implied.

In this work, we adopt three different cloud-size definitions: semi-major axis-based size $(R_{\rm{max}})$, area-based size $(R_S)$, and volume-based size $(R_V)$ as follows:

\begin{equation}
R_S=(\frac{S}{\pi})^{\frac{1}{2}}
\label{eqn:r_s}
\end{equation}

\begin{equation}
R_V=(\frac{3V}{4\pi})^{\frac{1}{3}}
\label{eqn:r_v}
\end{equation}
where $R_{\rm{max}}$ is the semi-major axes of the clouds calculated from Equation \ref{eqn:radius}, $S$ and $V$ are the area and volume of the clouds calculated from Equation \ref{eqn:area} and Equation \ref{eqn:volume} respectively.

The upper panel of Fig. \ref{fig:m-s} contains the mass-size relation of our sample clouds from three different size definitions. The fitting results are listed below:

\begin{equation}
M\propto 32.00\hspace{0.2em}{R_{\rm{max}}}^{1.77}
\label{eqn:m-rl}
\end{equation}

\begin{equation}
M\propto 20.59\hspace{0.2em}{R_S}^{2.04}
\label{eqn:m-rs}
\end{equation}

\begin{equation}
M\propto 14.41\hspace{0.2em}{R_V}^{2.29}
\label{eqn:m-rv}
\end{equation}
Our results reveal that all three size definitions follow a power-law mass-size relation, but each size definition exhibits a different exponent. We plot the results derived from \citet{Roman2010}, \citet{Miville2017}, and \citet{Chen2020} as a comparison. \citet{Roman2010} studied the physical properties of 580 molecular clouds using $^{12}\rm {CO}$ and $^{13}\rm {CO}$ emission line observation probed by the University of Massachusetts-Stony Brook (UMSB) and Galactic Ring surveys. \citet{Miville2017} used the $\rm {CO}$ surveys data from \citet{Dame2001} and identified a total of 8107 molecular clouds which cover the whole Galactic plane with their physical properties. \citet{Chen2020} recognized 567 molecular clouds within a $4\hspace{0.2em}\rm {kpc}$ from the Sun region in the 3D dust extinction map derived by \citet{Chen2019} and studied their physical properties. The lower panel of Fig. \ref{fig:m-s} shows the mass and size distribution of the clouds in this work. 

In Fig. \ref{fig:m-s}, the mass-size relations derived by our study and the previous three works have two major differences: First, for clouds with the same mass, our sample clouds have a larger size than those from previous studies. Second, different studies obtain different exponents of the mass-size relation $(M \sim R^{\beta})$. \citet{Chen2020} obtained the exponent $\beta=1.96$, which is quite close to our result on the area-based size definition of $\beta=2.04$. However, \citet{Roman2010} obtained the exponent of $\beta=2.36 \pm 0.04$ while \citet{Miville2017} derived $\beta=2.2 \pm 0.2$, which is similar to our result on the volume-based size definition of $\beta=2.29$. We propose that these discrepancies are partly attributed to our different definitions of cloud size. The three previous works calculated the cloud size based on the projected cloud area vertical to the line of sight. \citet{Roman2010} and \citet{Chen2020} calculated the solid angle $\Omega$ subtended by the clouds and cloud distance $d$ to us. They defined the cloud size as $R=\sqrt{S/\pi}$, where $S=\Omega d^2$. This is similar to our area-based size definition, but the distinction is that we adopt the largest area in the clouds (the $R_{\rm {max}}-R_{\rm {mid}}$ plane area, see Sec.\ref{Morphological parameter}). This may be why \citet{Chen2020} obtained an exponent quite close but smaller than our result on area-based size definition. \citet{Miville2017} used a size definition based on the moment of inertia. Still, their calculated radii are also based on the projected areas on the sky, and they defined the eigenvalues of the inertial matrix as cloud half-axes directly. In this case, our calculated size and area are larger for the same cloud than those in three previous works, especially for elongated clouds with low radial angles. We acknowledge different gas tracers (dust extinction vs. $\rm CO$ emission line), and different cloud identification methods (including different density thresholds for cloud identification) might affect the derived mass-size relations.


\begin{figure*}
\includegraphics[width=\textwidth]{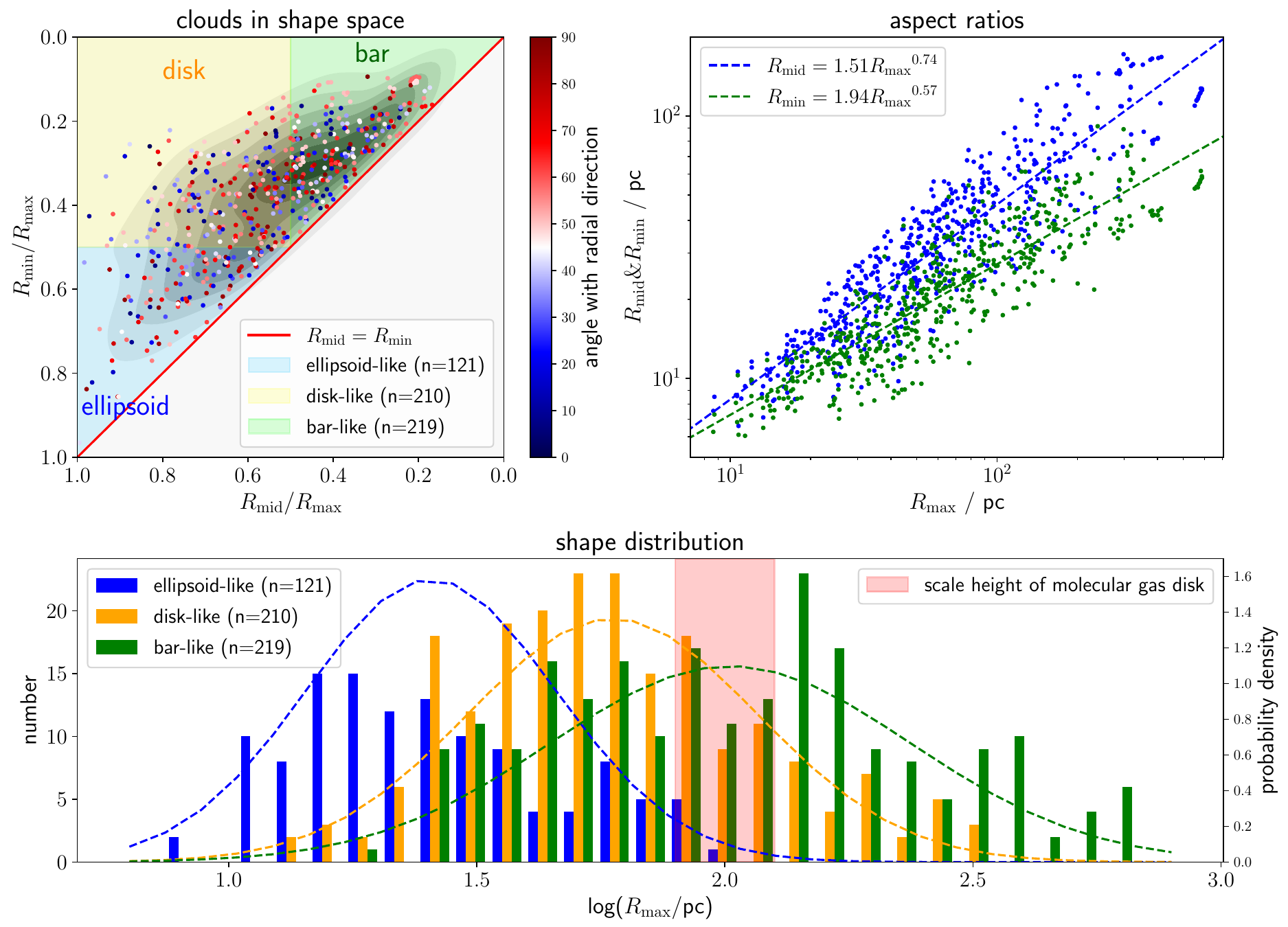}
\caption{\textbf{Shapes and aspect ratios of clouds.} (1) Upper left panel: clouds in shape space. The sample clouds are distinguished into ellipsoid-like clouds (number=121, blue region), disk-like clouds (number = 210, yellow region), and bar-like clouds (number = 219, green region) respectively. The color of the scatters corresponds to the angle between cloud orientations and the Galactic radial direction. The red solid line shows where $R_{\rm {mid}}=R_{\rm {min}}$. The background gray colormap shows the result of kernel density estimation. (2) Upper right panel: $R_{\rm{mid}}-R_{\rm{max}}$ and $R_{\rm{mid}}-R_{\rm{max}}$ relation of sample clouds. The fitting results are $R_{\rm{mid}} = 1.51{R_{\rm{max}}}^{0.74}$ (blue dashed line), and $R_{\rm{min}} = 1.94{R_{\rm{max}}}^{0.57}$ (green dashed line) respectively. (3) Lower panel: Distribution of clouds in different shapes as a function of cloud size. The red region represents the scale height of the molecular gas disk.}
\label{fig:ratio and shape}
\end{figure*}

\begin{figure*}
\includegraphics[width=\textwidth]{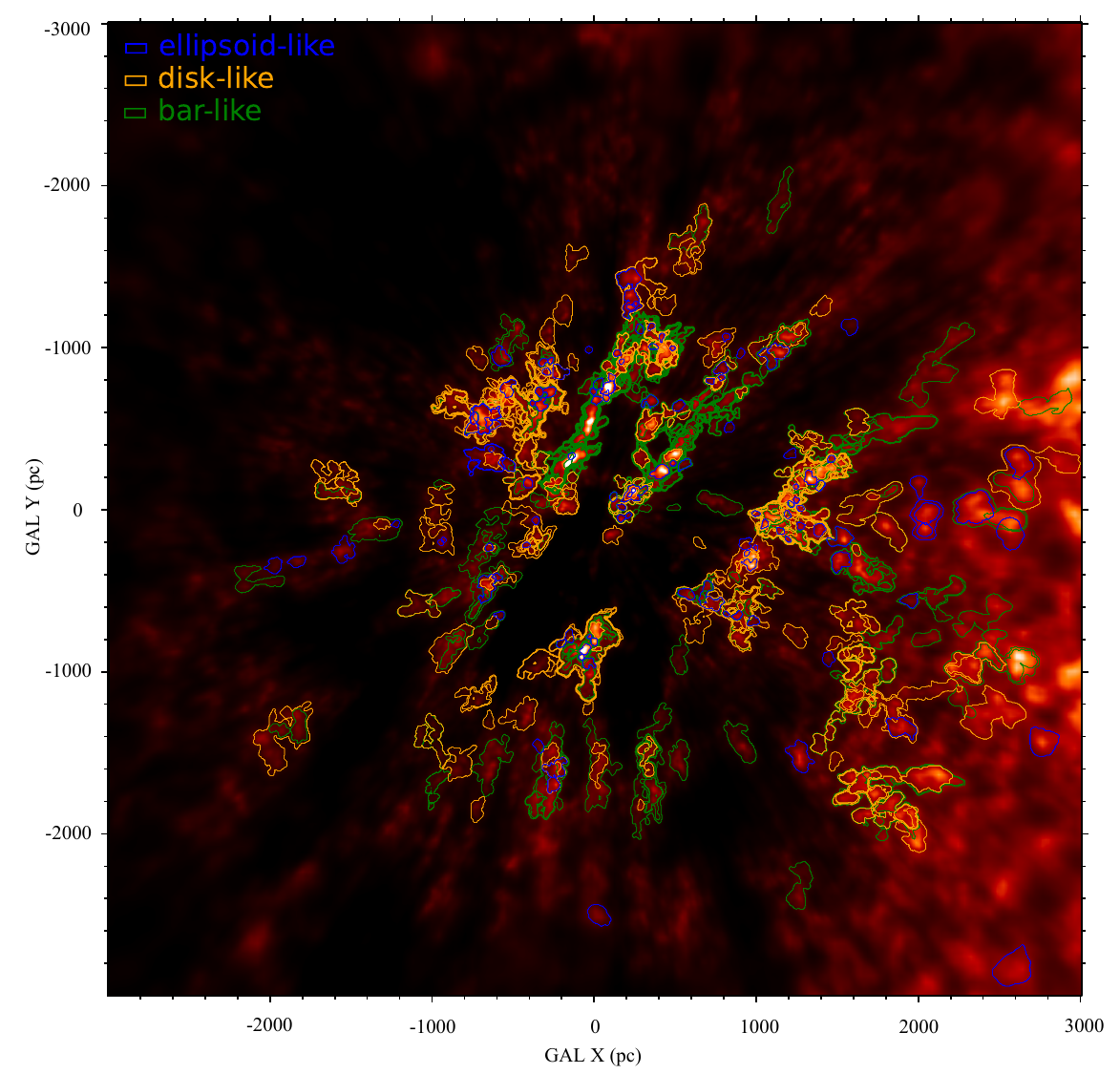}
\caption{\textbf{Contours of clouds in different shapes with the dust extinction map projected onto the x-y plane.} The clouds are in the shapes of ellipsoid-like (blue contour), disk-like (orange contour), and bar-like (green contour) respectively.}
\label{fig:contour}
\end{figure*}

\subsection{Density-Size Relation}

\subsubsection{Surface Density}

In Larson's third law, the mass surface density of molecular clouds is approximately constant and nearly independent of cloud size \citep{Larson1981}. We calculate the mass surface density of the sample clouds using Equation \ref{eqn: surface density}. Our fitting result of the surface density-size relation is shown in the left panel of Fig. \ref{fig:density} also exhibits similar properties as follows:

\begin{equation}
\Sigma \propto 6.75\hspace{0.2em}{R_{\rm{max}}}^{0.03}
\label{eqn:sd-r}
\end{equation}
where the exponent is almost equal to zero, indicating that the sample clouds have a constant mean mass surface density of $\sim 7\hspace{0.2em}\rm{M_{\odot}\hspace{0.2em}pc^{-2}}$.  

\citet{Krumholz2008, Krumholz2009} analysed the atomic-to-molecular transition based on the consideration of $\rm H_2$ formation, star formation, shielding and photodissociation. They conclude that the critical mass surface density of atomic-to-molecular transition is mainly determined by the metallicity. Based on this, \citet{Sternberg2014} derived the typical observed $\rm H_{\uppercase\expandafter{\romannumeral1}}/H_2$ transition surface density as follows:

\begin{equation}
\Sigma_{\rm crit} \approx \frac{12}{\phi_g Z}M_{\odot}\hspace{0.2em}\rm{pc}^{-2}
\label{eqn:critical sd}
\end{equation}
where $\phi_g$ is a factor of order unity depending on the dust grain absorption properties, and $Z=1$ corresponds to the solar metallicity. As our size definition provides a larger cloud area than previous studies, we expect that our measuring cloud surface densities should be smaller than the $\Sigma_{\rm{crit}}$. \citet{Lada2020} analyzed four molecular cloud catalogs from both dust extinction and $^{12}\rm{CO}$ emission line observation and concluded that outside of the molecular ring located between $4-7\hspace{0.2em}\rm{kpc}$ to the Galactic center, the observed Milky Way giant molecular clouds exhibit a constant mass surface density of $\Sigma_{\rm{GMC}}=35\pm8\hspace{0.2em}\rm{M_\odot}\hspace{0.2em}\rm{pc}^{-2}$.  Since they can not fully distinguish between different clouds using the velocity information alone, the cloud surface density can be overestimated. 

\subsubsection{Volume Density}

We calculate the cloud mass volume density via Equation \ref{eqn: volume density}. The result is shown in the right panel of Fig. \ref{fig:density}. \citet{Larson1981} derived the molecular cloud volume density-size relation of $\rho \sim R^{-1.10}$. However, our fitting result is as follows:

\begin{equation}
\rho \propto 2.60\hspace{0.2em}{R_{\rm{max}}}^{-0.55}
\label{eqn:vd-r}
\end{equation}
The exponent we obtain is smaller than one derived by  \citet{Larson1981}. This result is expected since we are studying structure at different scales. The rate of the decrease of gas density is slower on the kpc scale.


\subsection{Shapes and Aspect Ratios}
\label{shapes and aspect ratios}

We calculate the aspect ratios between the three axes of the molecular clouds of the sample. The results are presented in the upper right panel of Fig. \ref{fig:ratio and shape}. We find that the $R_{\rm{mid}}-R_{\rm{max}}$ and $R_{\rm{min}}-R_{\rm{max}}$ relation can be fitted by:

\begin{equation}
R_{\rm{mid}} = 1.51{R_{\rm{max}}}^{0.74}
\label{R_mid-R_max}
\end{equation}

\begin{equation}
R_{\rm{min}} = 1.94{R_{\rm{max}}}^{0.57}
\label{R_min-R_max}
\end{equation}
The exponents in Equation \ref{R_mid-R_max} and \ref{R_min-R_max} are smaller than 1, indicating that the aspect ratios between both $R_{\rm {mid}}$ and $R_{\rm {min}}$ with $R_{\rm {max}}$ decline with cloud size. This shows that the large clouds are more filamentary than small clouds. According to the value of $R_{\rm {mid}}/R_{\rm {max}}$ and $R_{\rm {min}}/R_{\rm {max}}$, we distinguish clouds into three morphological shapes: ellipsoid-like clouds $(R_{\rm {mid}}/R_{\rm {max}}>0.5, R_{\rm {min}}/R_{\rm {max}}>0.5)$, disk-like clouds $(R_{\rm {mid}}/R_{\rm {max}}>0.5, R_{\rm {min}}/R_{\rm {max}}<0.5)$, and bar-like clouds $(R_{\rm {mid}}/R_{\rm {max}}<0.5, R_{\rm {min}}/R_{\rm {max}}<0.5)$. Our results are shown in the upper left panel of Fig. \ref{fig:ratio and shape}.  The number of clouds in ellipsoid-like, disk-like, and bar-like shapes is 121, 210, and 219 respectively. The distribution of three shapes of clouds is shown in the lower panel of Fig. \ref{fig:ratio and shape} as a function of cloud size. As cloud size increases, their shapes gradually transition from ellipsoid-like to disk-like to bar-like. 

We plot the profiles of clouds in three shapes with the dust extinction map projected on the x-y plane in Fig. \ref{fig:contour}. We find that the ellipsoid-like clouds are in small and dense regions, while the largest structures exhibit a filamentary spatial shape.

We notice that even though our sample molecular clouds extend a wide range of aspect ratios in shape space, they still follow a power-law mass-size relation as well. This suggests that the mass-size relation $M \sim R^\beta$ is not dependent on the specific shapes of clouds, but widely applicable to molecular clouds with diverse morphology. However, the different definitions of cloud size could affect the exponents of mass-size \& volume density-size relations and the typical surface density of clouds we derive.


\subsection{Orientation}

\begin{figure*}
\includegraphics[width=\textwidth]{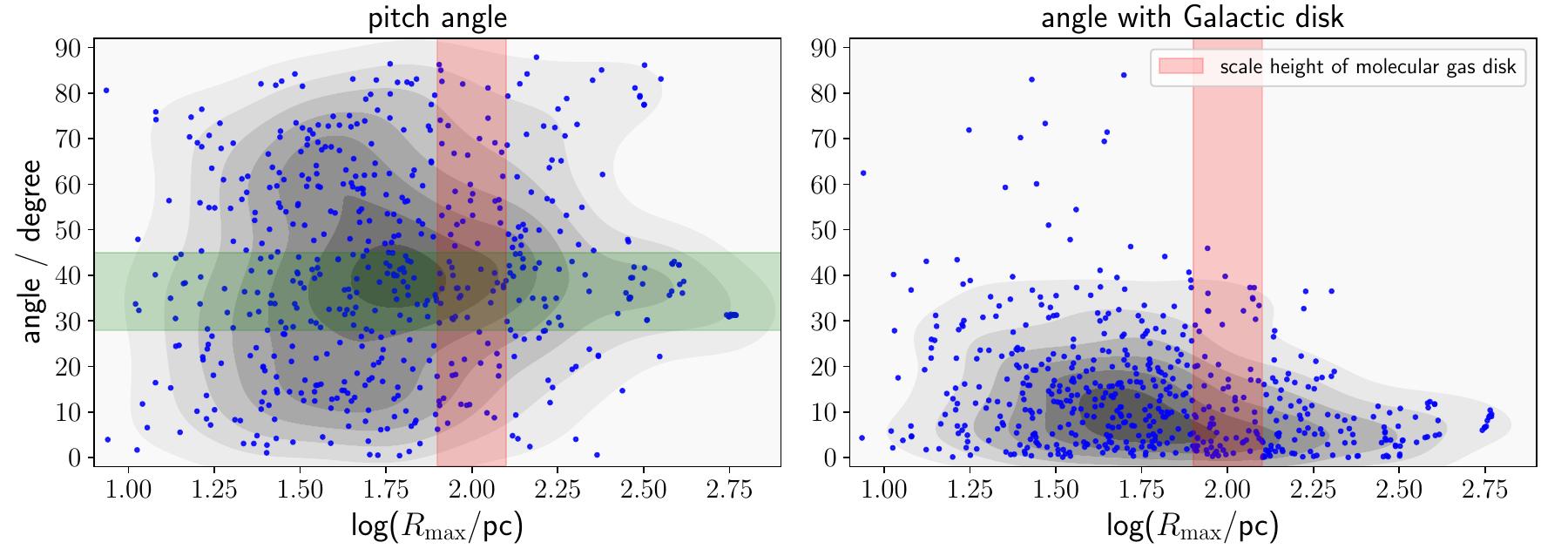}
\caption{\textbf{Orientations of clouds in the Milky Way.} (1) Left panel: pitch angle of clouds versus cloud size. (2) Right panel: angles between cloud orientations with the mid-plane of Galactic disk versus cloud size. The red region represents the scale height of the molecular gas disk. The background gray colormap shows the result of kernel density estimation.}
\label{fig:angle}
\end{figure*}

\subsubsection{Pitch Angle}
\label{radial direction}

We define the directions along the major axes ($\boldsymbol{R}_{\rm {max}}$) of clouds as their orientations in the Milky Way. The left panel of Fig. \ref{fig:angle} shows the pitch angle of clouds, where the angle is measured between cloud orientations with the Galactic tangential direction (we adopt the distance to the Galactic Center is $8.23\hspace{0.2em}\rm {kpc}$ from \citealt{Leung2023}). We find that for clouds over $100\hspace{0.2em}\rm {pc}$, their pitch angles tend to concentrate within a specific range of $28^{\circ}-45^{\circ}$.

We also calculate the extension of clouds along the Galactic radial direction. We use Equation \ref{eqn:rmean} to calculate the distance to the Galactic center of each cloud:

\begin{equation}
r_{\rm {mean}}=\frac{\sum_{i}r_i A_i}{\sum_{i}A_i} 
\label{eqn:rmean}
\end{equation}
where $r_i$ is the distance of each pixel within the cloud structure to the Galactic Center, and $A_i$ is the extinction value at each pixel. We use Equation \ref{eqn:deltar} to estimate the mass-weighted extension of clouds along the Galactic radial direction:

\begin{equation}
\Delta{r}=1.6\times\sqrt{\frac{\sum_{i}(r_i-r_{\rm {mean}})^2A_i}{\sum_{i}A_i}}
\label{eqn:deltar}
\end{equation}

\begin{figure*}
\includegraphics[width=\textwidth]{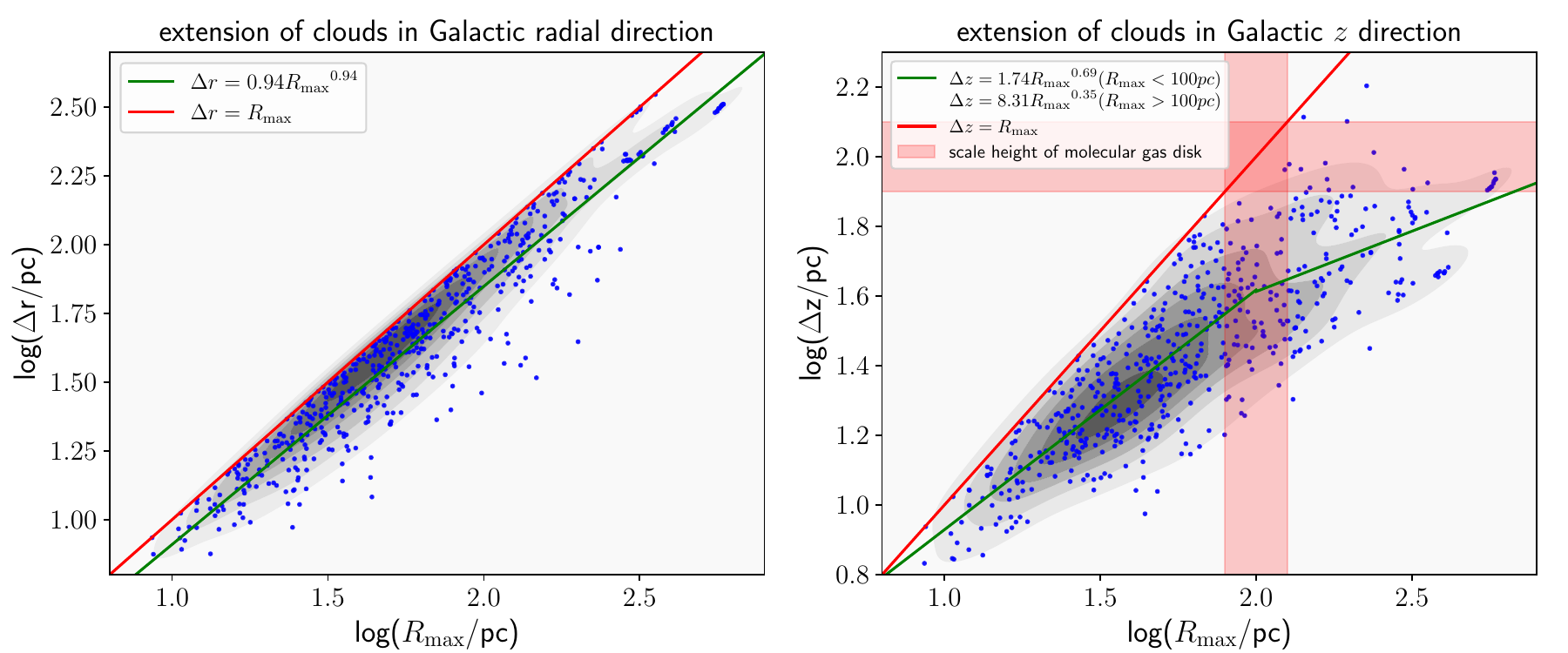}
\caption{\textbf{Extension of clouds along different directions.} (1) Left panel: extension of clouds along Galactic radial direction versus cloud size. The green solid line shows the fitting result $\Delta{r}=0.94{R_{\rm{max}}}^{0.94}$. The red solid line represents where $\Delta{r}={R_{\rm{max}}}$. (2) Right panel: extension of clouds along Galactic radial direction versus cloud size. The green solid line shows the fitting result $\Delta{z}=1.74{R_{\rm{max}}}^{0.69}\hspace{0.2em}(R_{\rm{max}}<100\rm pc)$ and $\Delta{z}=8.31{R_{\rm{max}}}^{0.35}\hspace{0.2em}(R_{\rm{max}}>100\rm pc)$. The red solid line represents where $\Delta{z}={R_{\rm{max}}}$. The red region represents the scale height of the molecular gas disk. The background gray colormap shows the result of kernel density estimation.}
\label{fig:delta z+r}
\end{figure*}

Our results of $\Delta{r}$ are shown in the left panel of Fig. \ref{fig:delta z+r} versus cloud size. The fitting result is:
\begin{equation}
\Delta{r}=0.94{R_{\rm{max}}}^{0.94}
\label{eqn:deltar-r}
\end{equation}
The exponent of the relation is quite close to 1, which implies that these clouds have a constant mean angle with the Galactic radial direction.

These results indicate that the galactic shear can affect the alignment of giant molecular clouds. The different rotative velocities at different Galactic radii can elongate giant molecular clouds within the Galactic disk along the tangential direction. Shear will reshape clouds into filamentary structures and induce their alignment within a specific range of pitch angles ($\sim 28^{\circ}-45^{\circ}$ in our case).

\subsubsection{Alignment with Galactic Disk}
\label{Alignment with Galactic disk}

We calculate the angle between the cloud orientations with the mid-plane of the Galactic disk. Our results are shown in the right panel of Fig. \ref{fig:angle}. We find that the cloud size of $\sim 100\hspace{0.2em}\rm {pc}$ - the scale height of the molecular gas disk - is a turning point. For clouds smaller than $100\hspace{0.2em}\rm {pc}$, their angles with the Galactic disk exhibit a range of distribution under $40^{\circ}$. But for clouds over $100\hspace{0.2em}\rm {pc}$, their angles with the Galactic disk concentrate more strongly under $20^{\circ}$. It suggests that large clouds tend to become parallel to the Galactic plane. This result provides another evidence of the effect of Galactic shear. It also indicates that large clouds are confined by the Galactic disk along the vertical direction.


\subsection{Vertical Confinement}

As already mentioned in Sec. \ref{Alignment with Galactic disk}, the large clouds stay parallel to the mid-plane of the Galactic disk. To further study the vertical confinement of the Galactic disk, we use the formula below to estimate the mass-weighted extension of clouds along the $z$ direction of the Galactic coordinate.

\begin{equation}
z_{\rm {mean}}=\frac{\sum_{i}z_i A_i}{\sum_{i}A_i} 
\label{eqn:zmean}
\end{equation}

\begin{equation}
\Delta{z}=1.6\times\sqrt{\frac{\sum_{i}(z_i-z_{\rm {mean}})^2A_i}{\sum_{i}A_i}}
\label{eqn:deltaz}
\end{equation}
where $z_{\rm {mean}}$ is the average distance to the Galactic plane of each cloud, $z_i$ is the distance of each pixel in cloud structures to the Galactic Center, and $A_i$ is the extinction value at each pixel. We use a piecewise power-law function to fit the $\Delta{z}-R_{\rm{max}}$ relation. The fitting results are shown in the right panel of Fig. \ref{fig:delta z+r} and as follows:
\begin{equation}
\Delta{z}=\left\{
\begin{aligned}
1.74{R_{\rm{max}}}^{0.69}\hspace{1em}(R_{\rm{max}}<100\rm pc) \\
8.31{R_{\rm{max}}}^{0.35}\hspace{1em}(R_{\rm{max}}>100\rm pc)
\end{aligned}
\right.
\label{eqn:deltaz-r}
\end{equation}

We find that when the size of the clouds is smaller than $100\hspace{0.2em}\rm {pc}$, their $\Delta{z}$ keeps increasing along with their size. But for cloud size larger than $100\hspace{0.2em}\rm {pc}$, the increasing rate of $\Delta{z}$ becomes slow and finally stops under $\sim 100\hspace{0.2em}\rm {pc}$ - the scale height of the molecular gas disk. This confinement is mostly caused by gravity from the stellar component, balanced by the cloud's turbulent motion. 


\section{Conclusion}

We identify 550 molecular clouds within a $6\times6\times0.8\hspace{0.2em}\rm kpc$ region of the solar vicinity based on the 3D extinction map through the Python program \textsc{Dendrogram}. We focus on the physical properties of the sample clouds, including their masses, three-axis size, and geometrical shapes. We also investigate the orientations of cloud structures in the Milky Way and calculate their extension along the Galactic radial direction and $z$ direction. Our main results are listed below:

\begin{itemize}
    \item {\bf Mass-size relation:} We use three different cloud size definitions: semi-major axis-based size $R_{\rm {max}}$, area-based size $R_S=(S/\pi)^{\frac{1}{2}}$ and volume-based size $R_V=({3V}/{4\pi})^{\frac{1}{3}}$. Based on them, we extend the mass-size relation of clouds in the Milky Way to the kiloparsec scale. The fitting results are $M\propto 32.00\hspace{0.2em}{R_{\rm{max}}}^{1.77}$, $M\propto 20.59\hspace{0.2em}{R_S}^{2.04}$ and $M\propto 14.41\hspace{0.2em}{R_V}^{2.29}$ respectively. 

    \item {\bf Cloud shape and evolution:} We categorize our sample molecular clouds into three different shapes - ellipsoid-like, disk-like, and bar-like - based on their aspect ratios ($R_{\rm {mid}}/R_{\rm {max}}$ and $R_{\rm {min}}/R_{\rm {max}}$). The small clouds are ellipsoid-like, medium-sized clouds tend to be disk-like, and large clouds are filament-like. 

    \item {\bf Orientation of the major-axis:} We investigate the pitch angles of clouds, where the angle is measured between cloud orientations with the Galactic tangential direction. We find that large cloud structures have a specific range of pitch angles of $\sim 28^{\circ}-45^{\circ}$.

    We further calculate the extension of the clouds along the Galactic radial direction $(\Delta r)$ and derive that its relation to the size of the clouds is $\Delta{r}=0.94{R_{\rm{max}}}^{0.94}$. The exponent here is close to 1, indicating that these clouds have a constant angle with the Galactic radial direction. This alignment is caused by the Galactic shear. 

    \item {\bf Vertical confinement:} We investigate the angles between cloud orientations with the mid-plane of the Galactic disk. Large clouds stay parallel to the mid-plane of the Galactic disk. 

\end{itemize}

We reveal a gradual increase in the importance of the Galactic processes on larger structures, where shear is responsible for a constant pitch angle of the large filament, and gravity from the weight of stars and gas can confine the gas to a height of about $100 \rm pc$ - the scale height of the molecular gas disk. The fact that structures tend to stay aligned with the mid-plane of the Milky Way disk is consistent with previous studies. This change is accompanied by a change from an ellipsoid-like cloud to a filament-like cloud. The results highlight the effect of shear and gravity at multiple scales in the star formation process.


\section*{Acknowledgement}
Yi-Heng Xie is supported by the Scientific Research and Innovation Project of Postgraduate Students in the Academic Degree of Yunnan University No. KC-23233846. Guang-Xing Li acknowledges support from NSFC grant No. 12273032 and 12033005. BQC is supported by the National Key R\&D Program of China No. 2019YFA0405500, National Natural Science Foundation of China 12173034 and 12322304, and the science research grants from the China Manned Space Project with NO. CMS-CSST-2021- A09, CMS-CSST-2021-A08, and CMS-CSST-2021-B03.


%

\vspace{5mm}
\facilities{Gaia, 2MASS}


\software{astropy \citep{astropy:2013, astropy:2018, astropy:2022}
          APLpy \citep{Robitaille2012}
          Dendrogram \citep{Rosolowsky2008}
          galpy \citep{Bovy2015}
          glue \citep{glue2015, glue2017}
          numpy \citep{numpy2020}
          scipy \citep{scipy2020}
          }




\bibliography{sample631}{}

\begin{thebibliography}{}
\expandafter\ifx\csname natexlab\endcsname\relax\def\natexlab#1{#1}\fi
\providecommand{\url}[1]{\href{#1}{#1}}
\providecommand{\dodoi}[1]{doi:~\href{http://doi.org/#1}{\nolinkurl{#1}}}
\providecommand{\doeprint}[1]{\href{http://ascl.net/#1}{\nolinkurl{http://ascl.net/#1}}}
\providecommand{\doarXiv}[1]{\href{https://arxiv.org/abs/#1}{\nolinkurl{https://arxiv.org/abs/#1}}}

\bibitem[{{Andr{\'e}} {et~al.}(2014){Andr{\'e}}, {Di Francesco}, {Ward-Thompson}, {Inutsuka}, {Pudritz}, \& {Pineda}}]{Andr2014}
{Andr{\'e}}, P., {Di Francesco}, J., {Ward-Thompson}, D., {et~al.} 2014, in Protostars and Planets VI, ed. H.~{Beuther}, R.~S. {Klessen}, C.~P. {Dullemond}, \& T.~{Henning}, 27--51, \dodoi{10.2458/azu_uapress_9780816531240-ch002}

\bibitem[{{Astropy Collaboration} {et~al.}(2013){Astropy Collaboration}, {Robitaille}, {Tollerud}, {Greenfield}, {Droettboom}, {Bray}, {Aldcroft}, {Davis}, {Ginsburg}, {Price-Whelan}, {Kerzendorf}, {Conley}, {Crighton}, {Barbary}, {Muna}, {Ferguson}, {Grollier}, {Parikh}, {Nair}, {Unther}, {Deil}, {Woillez}, {Conseil}, {Kramer}, {Turner}, {Singer}, {Fox}, {Weaver}, {Zabalza}, {Edwards}, {Azalee Bostroem}, {Burke}, {Casey}, {Crawford}, {Dencheva}, {Ely}, {Jenness}, {Labrie}, {Lim}, {Pierfederici}, {Pontzen}, {Ptak}, {Refsdal}, {Servillat}, \& {Streicher}}]{astropy:2013}
{Astropy Collaboration}, {Robitaille}, T.~P., {Tollerud}, E.~J., {et~al.} 2013, \aap, 558, A33, \dodoi{10.1051/0004-6361/201322068}

\bibitem[{{Astropy Collaboration} {et~al.}(2018){Astropy Collaboration}, {Price-Whelan}, {Sip{\H{o}}cz}, {G{\"u}nther}, {Lim}, {Crawford}, {Conseil}, {Shupe}, {Craig}, {Dencheva}, {Ginsburg}, {Vand erPlas}, {Bradley}, {P{\'e}rez-Su{\'a}rez}, {de Val-Borro}, {Aldcroft}, {Cruz}, {Robitaille}, {Tollerud}, {Ardelean}, {Babej}, {Bach}, {Bachetti}, {Bakanov}, {Bamford}, {Barentsen}, {Barmby}, {Baumbach}, {Berry}, {Biscani}, {Boquien}, {Bostroem}, {Bouma}, {Brammer}, {Bray}, {Breytenbach}, {Buddelmeijer}, {Burke}, {Calderone}, {Cano Rodr{\'\i}guez}, {Cara}, {Cardoso}, {Cheedella}, {Copin}, {Corrales}, {Crichton}, {D'Avella}, {Deil}, {Depagne}, {Dietrich}, {Donath}, {Droettboom}, {Earl}, {Erben}, {Fabbro}, {Ferreira}, {Finethy}, {Fox}, {Garrison}, {Gibbons}, {Goldstein}, {Gommers}, {Greco}, {Greenfield}, {Groener}, {Grollier}, {Hagen}, {Hirst}, {Homeier}, {Horton}, {Hosseinzadeh}, {Hu}, {Hunkeler}, {Ivezi{\'c}}, {Jain}, {Jenness}, {Kanarek}, {Kendrew}, {Kern}, {Kerzendorf}, {Khvalko}, {King}, {Kirkby}, {Kulkarni},
  {Kumar}, {Lee}, {Lenz}, {Littlefair}, {Ma}, {Macleod}, {Mastropietro}, {McCully}, {Montagnac}, {Morris}, {Mueller}, {Mumford}, {Muna}, {Murphy}, {Nelson}, {Nguyen}, {Ninan}, {N{\"o}the}, {Ogaz}, {Oh}, {Parejko}, {Parley}, {Pascual}, {Patil}, {Patil}, {Plunkett}, {Prochaska}, {Rastogi}, {Reddy Janga}, {Sabater}, {Sakurikar}, {Seifert}, {Sherbert}, {Sherwood-Taylor}, {Shih}, {Sick}, {Silbiger}, {Singanamalla}, {Singer}, {Sladen}, {Sooley}, {Sornarajah}, {Streicher}, {Teuben}, {Thomas}, {Tremblay}, {Turner}, {Terr{\'o}n}, {van Kerkwijk}, {de la Vega}, {Watkins}, {Weaver}, {Whitmore}, {Woillez}, {Zabalza}, \& {Astropy Contributors}}]{astropy:2018}
{Astropy Collaboration}, {Price-Whelan}, A.~M., {Sip{\H{o}}cz}, B.~M., {et~al.} 2018, \aj, 156, 123, \dodoi{10.3847/1538-3881/aabc4f}

\bibitem[{{Astropy Collaboration} {et~al.}(2022){Astropy Collaboration}, {Price-Whelan}, {Lim}, {Earl}, {Starkman}, {Bradley}, {Shupe}, {Patil}, {Corrales}, {Brasseur}, {N{"o}the}, {Donath}, {Tollerud}, {Morris}, {Ginsburg}, {Vaher}, {Weaver}, {Tocknell}, {Jamieson}, {van Kerkwijk}, {Robitaille}, {Merry}, {Bachetti}, {G{"u}nther}, {Aldcroft}, {Alvarado-Montes}, {Archibald}, {B{'o}di}, {Bapat}, {Barentsen}, {Baz{'a}n}, {Biswas}, {Boquien}, {Burke}, {Cara}, {Cara}, {Conroy}, {Conseil}, {Craig}, {Cross}, {Cruz}, {D'Eugenio}, {Dencheva}, {Devillepoix}, {Dietrich}, {Eigenbrot}, {Erben}, {Ferreira}, {Foreman-Mackey}, {Fox}, {Freij}, {Garg}, {Geda}, {Glattly}, {Gondhalekar}, {Gordon}, {Grant}, {Greenfield}, {Groener}, {Guest}, {Gurovich}, {Handberg}, {Hart}, {Hatfield-Dodds}, {Homeier}, {Hosseinzadeh}, {Jenness}, {Jones}, {Joseph}, {Kalmbach}, {Karamehmetoglu}, {Ka{l}uszy{'n}ski}, {Kelley}, {Kern}, {Kerzendorf}, {Koch}, {Kulumani}, {Lee}, {Ly}, {Ma}, {MacBride}, {Maljaars}, {Muna}, {Murphy}, {Norman}, {O'Steen},
  {Oman}, {Pacifici}, {Pascual}, {Pascual-Granado}, {Patil}, {Perren}, {Pickering}, {Rastogi}, {Roulston}, {Ryan}, {Rykoff}, {Sabater}, {Sakurikar}, {Salgado}, {Sanghi}, {Saunders}, {Savchenko}, {Schwardt}, {Seifert-Eckert}, {Shih}, {Jain}, {Shukla}, {Sick}, {Simpson}, {Singanamalla}, {Singer}, {Singhal}, {Sinha}, {Sip{H{o}}cz}, {Spitler}, {Stansby}, {Streicher}, {{{S}}umak}, {Swinbank}, {Taranu}, {Tewary}, {Tremblay}, {Val-Borro}, {Van Kooten}, {Vasovi{'c}}, {Verma}, {de Miranda Cardoso}, {Williams}, {Wilson}, {Winkel}, {Wood-Vasey}, {Xue}, {Yoachim}, {Zhang}, {Zonca}, \& {Astropy Project Contributors}}]{astropy:2022}
{Astropy Collaboration}, {Price-Whelan}, A.~M., {Lim}, P.~L., {et~al.} 2022, \apj, 935, 167, \dodoi{10.3847/1538-4357/ac7c74}

\bibitem[{{Ballesteros-Paredes} {et~al.}(2007){Ballesteros-Paredes}, {Klessen}, {Mac Low}, \& {Vazquez-Semadeni}}]{Ballesteros2007}
{Ballesteros-Paredes}, J., {Klessen}, R.~S., {Mac Low}, M.~M., \& {Vazquez-Semadeni}, E. 2007, in Protostars and Planets V, ed. B.~{Reipurth}, D.~{Jewitt}, \& K.~{Keil}, 63, \dodoi{10.48550/arXiv.astro-ph/0603357}

\bibitem[{{Bally} {et~al.}(1987){Bally}, {Langer}, {Stark}, \& {Wilson}}]{Bally1987}
{Bally}, J., {Langer}, W.~D., {Stark}, A.~A., \& {Wilson}, R.~W. 1987, \apjl, 312, L45, \dodoi{10.1086/184817}

\bibitem[{{Beaumont} {et~al.}(2015){Beaumont}, {Goodman}, \& {Greenfield}}]{glue2015}
{Beaumont}, C., {Goodman}, A., \& {Greenfield}, P. 2015, in Astronomical Society of the Pacific Conference Series, Vol. 495, Astronomical Data Analysis Software an Systems XXIV (ADASS XXIV), ed. A.~R. {Taylor} \& E.~{Rosolowsky}, 101

\bibitem[{{Bolatto} {et~al.}(2008){Bolatto}, {Leroy}, {Rosolowsky}, {Walter}, \& {Blitz}}]{Bolatto2008}
{Bolatto}, A.~D., {Leroy}, A.~K., {Rosolowsky}, E., {Walter}, F., \& {Blitz}, L. 2008, \apj, 686, 948, \dodoi{10.1086/591513}

\bibitem[{{Bovy}(2015)}]{Bovy2015}
{Bovy}, J. 2015, \apjs, 216, 29, \dodoi{10.1088/0067-0049/216/2/29}

\bibitem[{{Chen} {et~al.}(2015){Chen}, {Liu}, {Yuan}, {Huang}, \& {Xiang}}]{Chen2015}
{Chen}, B.~Q., {Liu}, X.~W., {Yuan}, H.~B., {Huang}, Y., \& {Xiang}, M.~S. 2015, \mnras, 448, 2187, \dodoi{10.1093/mnras/stv103}

\bibitem[{{Chen} {et~al.}(2019){Chen}, {Huang}, {Yuan}, {Wang}, {Fan}, {Xiang}, {Zhang}, {Tian}, \& {Liu}}]{Chen2019}
{Chen}, B.~Q., {Huang}, Y., {Yuan}, H.~B., {et~al.} 2019, \mnras, 483, 4277, \dodoi{10.1093/mnras/sty3341}

\bibitem[{{Chen} {et~al.}(2020){Chen}, {Li}, {Yuan}, {Huang}, {Tian}, {Wang}, {Zhang}, {Wang}, \& {Liu}}]{Chen2020}
{Chen}, B.~Q., {Li}, G.~X., {Yuan}, H.~B., {et~al.} 2020, \mnras, 493, 351, \dodoi{10.1093/mnras/staa235}

\bibitem[{{Crutcher}(2012)}]{Richard2012}
{Crutcher}, R.~M. 2012, \araa, 50, 29, \dodoi{10.1146/annurev-astro-081811-125514}

\bibitem[{{Dame} {et~al.}(2001){Dame}, {Hartmann}, \& {Thaddeus}}]{Dame2001}
{Dame}, T.~M., {Hartmann}, D., \& {Thaddeus}, P. 2001, \apj, 547, 792, \dodoi{10.1086/318388}

\bibitem[{{Duarte-Cabral} \& {Dobbs}(2017)}]{Duarte2017}
{Duarte-Cabral}, A., \& {Dobbs}, C.~L. 2017, \mnras, 470, 4261, \dodoi{10.1093/mnras/stx1524}

\bibitem[{{Elmegreen} \& {Scalo}(2004)}]{Elmegreen2004}
{Elmegreen}, B.~G., \& {Scalo}, J. 2004, \araa, 42, 211, \dodoi{10.1146/annurev.astro.41.011802.094859}

\bibitem[{{Gaia Collaboration} {et~al.}(2021){Gaia Collaboration}, {Brown}, {Vallenari}, {Prusti}, {de Bruijne}, {Babusiaux}, {Biermann}, {Creevey}, {Evans}, {Eyer}, {Hutton}, {Jansen}, {Jordi}, {Klioner}, {Lammers}, {Lindegren}, {Luri}, {Mignard}, {Panem}, {Pourbaix}, {Randich}, {Sartoretti}, {Soubiran}, {Walton}, {Arenou}, {Bailer-Jones}, {Bastian}, {Cropper}, {Drimmel}, {Katz}, {Lattanzi}, {van Leeuwen}, {Bakker}, {Cacciari}, {Casta{\~n}eda}, {De Angeli}, {Ducourant}, {Fabricius}, {Fouesneau}, {Fr{\'e}mat}, {Guerra}, {Guerrier}, {Guiraud}, {Jean-Antoine Piccolo}, {Masana}, {Messineo}, {Mowlavi}, {Nicolas}, {Nienartowicz}, {Pailler}, {Panuzzo}, {Riclet}, {Roux}, {Seabroke}, {Sordo}, {Tanga}, {Th{\'e}venin}, {Gracia-Abril}, {Portell}, {Teyssier}, {Altmann}, {Andrae}, {Bellas-Velidis}, {Benson}, {Berthier}, {Blomme}, {Brugaletta}, {Burgess}, {Busso}, {Carry}, {Cellino}, {Cheek}, {Clementini}, {Damerdji}, {Davidson}, {Delchambre}, {Dell'Oro}, {Fern{\'a}ndez-Hern{\'a}ndez}, {Galluccio}, {Garc{\'\i}a-Lario},
  {Garcia-Reinaldos}, {Gonz{\'a}lez-N{\'u}{\~n}ez}, {Gosset}, {Haigron}, {Halbwachs}, {Hambly}, {Harrison}, {Hatzidimitriou}, {Heiter}, {Hern{\'a}ndez}, {Hestroffer}, {Hodgkin}, {Holl}, {Jan{\ss}en}, {Jevardat de Fombelle}, {Jordan}, {Krone-Martins}, {Lanzafame}, {L{\"o}ffler}, {Lorca}, {Manteiga}, {Marchal}, {Marrese}, {Moitinho}, {Mora}, {Muinonen}, {Osborne}, {Pancino}, {Pauwels}, {Petit}, {Recio-Blanco}, {Richards}, {Riello}, {Rimoldini}, {Robin}, {Roegiers}, {Rybizki}, {Sarro}, {Siopis}, {Smith}, {Sozzetti}, {Ulla}, {Utrilla}, {van Leeuwen}, {van Reeven}, {Abbas}, {Abreu Aramburu}, {Accart}, {Aerts}, {Aguado}, {Ajaj}, {Altavilla}, {{\'A}lvarez}, {{\'A}lvarez Cid-Fuentes}, {Alves}, {Anderson}, {Anglada Varela}, {Antoja}, {Audard}, {Baines}, {Baker}, {Balaguer-N{\'u}{\~n}ez}, {Balbinot}, {Balog}, {Barache}, {Barbato}, {Barros}, {Barstow}, {Bartolom{\'e}}, {Bassilana}, {Bauchet}, {Baudesson-Stella}, {Becciani}, {Bellazzini}, {Bernet}, {Bertone}, {Bianchi}, {Blanco-Cuaresma}, {Boch}, {Bombrun}, {Bossini},
  {Bouquillon}, {Bragaglia}, {Bramante}, {Breedt}, {Bressan}, {Brouillet}, {Bucciarelli}, {Burlacu}, {Busonero}, {Butkevich}, {Buzzi}, {Caffau}, {Cancelliere}, {C{\'a}novas}, {Cantat-Gaudin}, {Carballo}, {Carlucci}, {Carnerero}, {Carrasco}, {Casamiquela}, {Castellani}, {Castro-Ginard}, {Castro Sampol}, {Chaoul}, {Charlot}, {Chemin}, {Chiavassa}, {Cioni}, {Comoretto}, {Cooper}, {Cornez}, {Cowell}, {Crifo}, {Crosta}, {Crowley}, {Dafonte}, {Dapergolas}, {David}, {David}, {de Laverny}, {De Luise}, {De March}, {De Ridder}, {de Souza}, {de Teodoro}, {de Torres}, {del Peloso}, {del Pozo}, {Delbo}, {Delgado}, {Delgado}, {Delisle}, {Di Matteo}, {Diakite}, {Diener}, {Distefano}, {Dolding}, {Eappachen}, {Edvardsson}, {Enke}, {Esquej}, {Fabre}, {Fabrizio}, {Faigler}, {Fedorets}, {Fernique}, {Fienga}, {Figueras}, {Fouron}, {Fragkoudi}, {Fraile}, {Franke}, {Gai}, {Garabato}, {Garcia-Gutierrez}, {Garc{\'\i}a-Torres}, {Garofalo}, {Gavras}, {Gerlach}, {Geyer}, {Giacobbe}, {Gilmore}, {Girona}, {Giuffrida}, {Gomel}, {Gomez},
  {Gonzalez-Santamaria}, {Gonz{\'a}lez-Vidal}, {Granvik}, {Guti{\'e}rrez-S{\'a}nchez}, {Guy}, {Hauser}, {Haywood}, {Helmi}, {Hidalgo}, {Hilger}, {H{\l}adczuk}, {Hobbs}, {Holland}, {Huckle}, {Jasniewicz}, {Jonker}, {Juaristi Campillo}, {Julbe}, {Karbevska}, {Kervella}, {Khanna}, {Kochoska}, {Kontizas}, {Kordopatis}, {Korn}, {Kostrzewa-Rutkowska}, {Kruszy{\'n}ska}, {Lambert}, {Lanza}, {Lasne}, {Le Campion}, {Le Fustec}, {Lebreton}, {Lebzelter}, {Leccia}, {Leclerc}, {Lecoeur-Taibi}, {Liao}, {Licata}, {Lindstr{\o}m}, {Lister}, {Livanou}, {Lobel}, {Madrero Pardo}, {Managau}, {Mann}, {Marchant}, {Marconi}, {Marcos Santos}, {Marinoni}, {Marocco}, {Marshall}, {Martin Polo}, {Mart{\'\i}n-Fleitas}, {Masip}, {Massari}, {Mastrobuono-Battisti}, {Mazeh}, {McMillan}, {Messina}, {Michalik}, {Millar}, {Mints}, {Molina}, {Molinaro}, {Moln{\'a}r}, {Montegriffo}, {Mor}, {Morbidelli}, {Morel}, {Morris}, {Mulone}, {Munoz}, {Muraveva}, {Murphy}, {Musella}, {Noval}, {Ord{\'e}novic}, {Orr{\`u}}, {Osinde}, {Pagani}, {Pagano},
  {Palaversa}, {Palicio}, {Panahi}, {Pawlak}, {Pe{\~n}alosa Esteller}, {Penttil{\"a}}, {Piersimoni}, {Pineau}, {Plachy}, {Plum}, {Poggio}, {Poretti}, {Poujoulet}, {Pr{\v{s}}a}, {Pulone}, {Racero}, {Ragaini}, {Rainer}, {Raiteri}, {Rambaux}, {Ramos}, {Ramos-Lerate}, {Re Fiorentin}, {Regibo}, {Reyl{\'e}}, {Ripepi}, {Riva}, {Rixon}, {Robichon}, {Robin}, {Roelens}, {Rohrbasser}, {Romero-G{\'o}mez}, {Rowell}, {Royer}, {Rybicki}, {Sadowski}, {Sagrist{\`a} Sell{\'e}s}, {Sahlmann}, {Salgado}, {Salguero}, {Samaras}, {Sanchez Gimenez}, {Sanna}, {Santove{\~n}a}, {Sarasso}, {Schultheis}, {Sciacca}, {Segol}, {Segovia}, {S{\'e}gransan}, {Semeux}, {Shahaf}, {Siddiqui}, {Siebert}, {Siltala}, {Slezak}, {Smart}, {Solano}, {Solitro}, {Souami}, {Souchay}, {Spagna}, {Spoto}, {Steele}, {Steidelm{\"u}ller}, {Stephenson}, {S{\"u}veges}, {Szabados}, {Szegedi-Elek}, {Taris}, {Tauran}, {Taylor}, {Teixeira}, {Thuillot}, {Tonello}, {Torra}, {Torra}, {Turon}, {Unger}, {Vaillant}, {van Dillen}, {Vanel}, {Vecchiato}, {Viala}, {Vicente},
  {Voutsinas}, {Weiler}, {Wevers}, {Wyrzykowski}, {Yoldas}, {Yvard}, {Zhao}, {Zorec}, {Zucker}, {Zurbach}, \& {Zwitter}}]{Gaia2021}
{Gaia Collaboration}, {Brown}, A.~G.~A., {Vallenari}, A., {et~al.} 2021, \aap, 649, A1, \dodoi{10.1051/0004-6361/202039657}

\bibitem[{{Goldsmith} {et~al.}(2008){Goldsmith}, {Heyer}, {Narayanan}, {Snell}, {Li}, \& {Brunt}}]{Goldsmith2008}
{Goldsmith}, P.~F., {Heyer}, M., {Narayanan}, G., {et~al.} 2008, \apj, 680, 428, \dodoi{10.1086/587166}

\bibitem[{{Hacar} {et~al.}(2023){Hacar}, {Clark}, {Heitsch}, {Kainulainen}, {Panopoulou}, {Seifried}, \& {Smith}}]{Hacar2023}
{Hacar}, A., {Clark}, S.~E., {Heitsch}, F., {et~al.} 2023, in Astronomical Society of the Pacific Conference Series, Vol. 534, Protostars and Planets VII, ed. S.~{Inutsuka}, Y.~{Aikawa}, T.~{Muto}, K.~{Tomida}, \& M.~{Tamura}, 153, \dodoi{10.48550/arXiv.2203.09562}

\bibitem[{Harris {et~al.}(2020)Harris, Millman, van~der Walt, Gommers, Virtanen, Cournapeau, Wieser, Taylor, Berg, Smith, Kern, Picus, Hoyer, van Kerkwijk, Brett, Haldane, del R{\'{i}}o, Wiebe, Peterson, G{\'{e}}rard-Marchant, Sheppard, Reddy, Weckesser, Abbasi, Gohlke, \& Oliphant}]{numpy2020}
Harris, C.~R., Millman, K.~J., van~der Walt, S.~J., {et~al.} 2020, Nature, 585, 357, \dodoi{10.1038/s41586-020-2649-2}

\bibitem[{{Jeffreson} \& {Kruijssen}(2018)}]{Jeffreson2018}
{Jeffreson}, S. M.~R., \& {Kruijssen}, J.~M.~D. 2018, \mnras, 476, 3688, \dodoi{10.1093/mnras/sty594}

\bibitem[{{Krumholz} {et~al.}(2008){Krumholz}, {McKee}, \& {Tumlinson}}]{Krumholz2008}
{Krumholz}, M.~R., {McKee}, C.~F., \& {Tumlinson}, J. 2008, \apj, 689, 865, \dodoi{10.1086/592490}

\bibitem[{{Krumholz} {et~al.}(2009){Krumholz}, {McKee}, \& {Tumlinson}}]{Krumholz2009}
---. 2009, \apj, 693, 216, \dodoi{10.1088/0004-637X/693/1/216}

\bibitem[{{Lada} \& {Dame}(2020)}]{Lada2020}
{Lada}, C.~J., \& {Dame}, T.~M. 2020, \apj, 898, 3, \dodoi{10.3847/1538-4357/ab9bfb}

\bibitem[{{Lallement} {et~al.}(2019){Lallement}, {Babusiaux}, {Vergely}, {Katz}, {Arenou}, {Valette}, {Hottier}, \& {Capitanio}}]{Lallement2019}
{Lallement}, R., {Babusiaux}, C., {Vergely}, J.~L., {et~al.} 2019, \aap, 625, A135, \dodoi{10.1051/0004-6361/201834695}

\bibitem[{{Larson}(1981)}]{Larson1981}
{Larson}, R.~B. 1981, \mnras, 194, 809, \dodoi{10.1093/mnras/194.4.809}

\bibitem[{{Lee} \& {Hennebelle}(2018)}]{Lee2018}
{Lee}, Y.-N., \& {Hennebelle}, P. 2018, \aap, 611, A89, \dodoi{10.1051/0004-6361/201731523}

\bibitem[{{Leung} {et~al.}(2023){Leung}, {Bovy}, {Mackereth}, {Hunt}, {Lane}, \& {Wilson}}]{Leung2023}
{Leung}, H.~W., {Bovy}, J., {Mackereth}, J.~T., {et~al.} 2023, \mnras, 519, 948, \dodoi{10.1093/mnras/stac3529}

\bibitem[{{Li}(2017)}]{Li2017}
{Li}, G.-X. 2017, \mnras, 465, 667, \dodoi{10.1093/mnras/stw2707}

\bibitem[{{Li}(2024)}]{Li2024}
---. 2024, \mnras, 528, L52, \dodoi{10.1093/mnrasl/slad149}

\bibitem[{{Li} {et~al.}(2016){Li}, {Urquhart}, {Leurini}, {Csengeri}, {Wyrowski}, {Menten}, \& {Schuller}}]{Li2016}
{Li}, G.-X., {Urquhart}, J.~S., {Leurini}, S., {et~al.} 2016, \aap, 591, A5, \dodoi{10.1051/0004-6361/201527468}

\bibitem[{{Li} {et~al.}(2013){Li}, {Wyrowski}, {Menten}, \& {Belloche}}]{Li2013}
{Li}, G.-X., {Wyrowski}, F., {Menten}, K., \& {Belloche}, A. 2013, \aap, 559, A34, \dodoi{10.1051/0004-6361/201322411}

\bibitem[{{Li} {et~al.}(2022){Li}, {Zhou}, \& {Chen}}]{Li2022}
{Li}, G.-X., {Zhou}, J.-X., \& {Chen}, B.-Q. 2022, \mnras, 516, L35, \dodoi{10.1093/mnrasl/slac076}

\bibitem[{{Li} {et~al.}(2014){Li}, {Goodman}, {Sridharan}, {Houde}, {Li}, {Novak}, \& {Tang}}]{LiHB2014}
{Li}, H.~B., {Goodman}, A., {Sridharan}, T.~K., {et~al.} 2014, in Protostars and Planets VI, ed. H.~{Beuther}, R.~S. {Klessen}, C.~P. {Dullemond}, \& T.~{Henning}, 101--123, \dodoi{10.2458/azu_uapress_9780816531240-ch005}

\bibitem[{{Liu} {et~al.}(2021){Liu}, {Bureau}, {Blitz}, {Davis}, {Onishi}, {Smith}, {North}, \& {Iguchi}}]{Liu2021}
{Liu}, L., {Bureau}, M., {Blitz}, L., {et~al.} 2021, \mnras, 505, 4048, \dodoi{10.1093/mnras/stab1537}

\bibitem[{{Lombardi} {et~al.}(2011){Lombardi}, {Alves}, \& {Lada}}]{Lombardi2011}
{Lombardi}, M., {Alves}, J., \& {Lada}, C.~J. 2011, \aap, 535, A16, \dodoi{10.1051/0004-6361/201116915}

\bibitem[{{Miville-Desch{\^e}nes} {et~al.}(2017){Miville-Desch{\^e}nes}, {Murray}, \& {Lee}}]{Miville2017}
{Miville-Desch{\^e}nes}, M.-A., {Murray}, N., \& {Lee}, E.~J. 2017, \apj, 834, 57, \dodoi{10.3847/1538-4357/834/1/57}

\bibitem[{{Neralwar} {et~al.}(2022{\natexlab{a}}){Neralwar}, {Colombo}, {Duarte-Cabral}, {Urquhart}, {Mattern}, {Wyrowski}, {Menten}, {Barnes}, {S{\'a}nchez-Monge}, {Beuther}, {Rigby}, {Mazumdar}, {Eden}, {Csengeri}, {Dobbs}, {Veena}, {Neupane}, {Henning}, {Schuller}, {Leurini}, {Wienen}, {Yang}, {Ragan}, {Medina}, \& {Nguyen-Luong}}]{Neralwar2022a}
{Neralwar}, K.~R., {Colombo}, D., {Duarte-Cabral}, A., {et~al.} 2022{\natexlab{a}}, \aap, 663, A56, \dodoi{10.1051/0004-6361/202142428}

\bibitem[{{Neralwar} {et~al.}(2022{\natexlab{b}}){Neralwar}, {Colombo}, {Duarte-Cabral}, {Urquhart}, {Mattern}, {Wyrowski}, {Menten}, {Barnes}, {S{\'a}nchez-Monge}, {Rigby}, {Mazumdar}, {Eden}, {Csengeri}, {Dobbs}, {Veena}, {Neupane}, {Henning}, {Schuller}, {Leurini}, {Wienen}, {Yang}, {Ragan}, {Medina}, \& {Nguyen-Luong}}]{Neralwar2022b}
---. 2022{\natexlab{b}}, \aap, 664, A84, \dodoi{10.1051/0004-6361/202142513}

\bibitem[{{Robitaille} {et~al.}(2017){Robitaille}, {Beaumont}, {Qian}, {Borkin}, \& {Goodman}}]{glue2017}
{Robitaille}, T., {Beaumont}, C., {Qian}, P., {Borkin}, M., \& {Goodman}, A. 2017, {glueviz v0.13.1: multidimensional data exploration}, 0.13.1,  Zenodo, \dodoi{10.5281/zenodo.1237692}

\bibitem[{{Robitaille} \& {Bressert}(2012)}]{Robitaille2012}
{Robitaille}, T., \& {Bressert}, E. 2012, {APLpy: Astronomical Plotting Library in Python}, Astrophysics Source Code Library, record ascl:1208.017

\bibitem[{{Roman-Duval} {et~al.}(2010){Roman-Duval}, {Jackson}, {Heyer}, {Rathborne}, \& {Simon}}]{Roman2010}
{Roman-Duval}, J., {Jackson}, J.~M., {Heyer}, M., {Rathborne}, J., \& {Simon}, R. 2010, \apj, 723, 492, \dodoi{10.1088/0004-637X/723/1/492}

\bibitem[{{Rosolowsky} {et~al.}(2008){Rosolowsky}, {Pineda}, {Kauffmann}, \& {Goodman}}]{Rosolowsky2008}
{Rosolowsky}, E.~W., {Pineda}, J.~E., {Kauffmann}, J., \& {Goodman}, A.~A. 2008, \apj, 679, 1338, \dodoi{10.1086/587685}

\bibitem[{{Scalo} \& {Elmegreen}(2004)}]{Scalo2004}
{Scalo}, J., \& {Elmegreen}, B.~G. 2004, \araa, 42, 275, \dodoi{10.1146/annurev.astro.42.120403.143327}

\bibitem[{{Skrutskie} {et~al.}(2006){Skrutskie}, {Cutri}, {Stiening}, {Weinberg}, {Schneider}, {Carpenter}, {Beichman}, {Capps}, {Chester}, {Elias}, {Huchra}, {Liebert}, {Lonsdale}, {Monet}, {Price}, {Seitzer}, {Jarrett}, {Kirkpatrick}, {Gizis}, {Howard}, {Evans}, {Fowler}, {Fullmer}, {Hurt}, {Light}, {Kopan}, {Marsh}, {McCallon}, {Tam}, {Van Dyk}, \& {Wheelock}}]{2MASS2006}
{Skrutskie}, M.~F., {Cutri}, R.~M., {Stiening}, R., {et~al.} 2006, \aj, 131, 1163, \dodoi{10.1086/498708}

\bibitem[{{Smith} {et~al.}(2014){Smith}, {Glover}, {Clark}, {Klessen}, \& {Springel}}]{Smith2014}
{Smith}, R.~J., {Glover}, S. C.~O., {Clark}, P.~C., {Klessen}, R.~S., \& {Springel}, V. 2014, \mnras, 441, 1628, \dodoi{10.1093/mnras/stu616}

\bibitem[{{Solomon} {et~al.}(1987){Solomon}, {Rivolo}, {Barrett}, \& {Yahil}}]{Solomon1987}
{Solomon}, P.~M., {Rivolo}, A.~R., {Barrett}, J., \& {Yahil}, A. 1987, \apj, 319, 730, \dodoi{10.1086/165493}

\bibitem[{{Sternberg} {et~al.}(2014){Sternberg}, {Le Petit}, {Roueff}, \& {Le Bourlot}}]{Sternberg2014}
{Sternberg}, A., {Le Petit}, F., {Roueff}, E., \& {Le Bourlot}, J. 2014, \apj, 790, 10, \dodoi{10.1088/0004-637X/790/1/10}

\bibitem[{{Vergely} {et~al.}(2022){Vergely}, {Lallement}, \& {Cox}}]{Vergely2022}
{Vergely}, J.~L., {Lallement}, R., \& {Cox}, N.~L.~J. 2022, \aap, 664, A174, \dodoi{10.1051/0004-6361/202243319}

\bibitem[{{Vergely} {et~al.}(2010){Vergely}, {Valette}, {Lallement}, \& {Raimond}}]{Vergely2010}
{Vergely}, J.~L., {Valette}, B., {Lallement}, R., \& {Raimond}, S. 2010, \aap, 518, A31, \dodoi{10.1051/0004-6361/200913962}

\bibitem[{Virtanen {et~al.}(2020)Virtanen, Gommers, Oliphant, Haberland, Reddy, Cournapeau, Burovski, Peterson, Weckesser, Bright, {van der Walt}, Brett, Wilson, Millman, Mayorov, Nelson, Jones, Kern, Larson, Carey, Polat, Feng, Moore, {VanderPlas}, Laxalde, Perktold, Cimrman, Henriksen, Quintero, Harris, Archibald, Ribeiro, Pedregosa, {van Mulbregt}, \& {SciPy 1.0 Contributors}}]{scipy2020}
Virtanen, P., Gommers, R., Oliphant, T.~E., {et~al.} 2020, Nature Methods, 17, 261, \dodoi{10.1038/s41592-019-0686-2}

\bibitem[{{Wang} {et~al.}(2015){Wang}, {Testi}, {Ginsburg}, {Walmsley}, {Molinari}, \& {Schisano}}]{Wang2015}
{Wang}, K., {Testi}, L., {Ginsburg}, A., {et~al.} 2015, \mnras, 450, 4043, \dodoi{10.1093/mnras/stv735}

\bibitem[{{Yuan} {et~al.}(2021){Yuan}, {Yang}, {Du}, {Liu}, {Zhang}, {Lin}, {Sun}, {Yan}, {Ma}, {Su}, {Sun}, \& {Zhou}}]{Yuan2021}
{Yuan}, L., {Yang}, J., {Du}, F., {et~al.} 2021, \apjs, 257, 51, \dodoi{10.3847/1538-4365/ac242a}

\bibitem[{{Zucker} {et~al.}(2015){Zucker}, {Battersby}, \& {Goodman}}]{Zucker2015}
{Zucker}, C., {Battersby}, C., \& {Goodman}, A. 2015, \apj, 815, 23, \dodoi{10.1088/0004-637X/815/1/23}

\end{thebibliography}
\bibliographystyle{aasjournal}



\end{document}